\documentclass[titlepage,12pt]{article}
\usepackage{amsfonts}
\usepackage{amsmath}
\usepackage[dvips]{graphicx}
\usepackage{subfigure}
\usepackage[font=small,format=plain,labelfont=bf,up,textfont=it,up]{caption}

\begin{document}

\title{Stationary states of fermions in a sign potential\\
with a mixed vector-scalar coupling\thanks{%
Annals of Physics 340/1 (2014), pp. 1-12 DOI: 10.1016/j.aop.2013.10.008}}
\date{}
\author{W.M. Castilho\thanks{%
E-mail address: castilho.w@gmail.com (W.M. Castilho)} and A.S. de Castro%
\thanks{%
E-mail address: castro@pq.cnpq.br (A.S. de Castro)} \\
\\
UNESP - Campus de Guaratinguet\'{a}\\
Departamento de F\'{\i}sica e Qu\'{\i}mica\\
12516-410 Guaratinguet\'{a} SP - Brasil }
\date{}
\maketitle

\begin{abstract}
The scattering of a fermion in the background of a sign potential is
considered with a general mixing of vector and scalar Lorentz structures
with the scalar coupling stronger than or equal to the vector coupling under
the Sturm-Liouville perspective. When the vector coupling and the scalar
coupling have different magnitudes, an isolated solution shows that the
fermion under a strong potential can be trapped in a highly localized region
without manifestation of Klein's paradox. It is also shown that the lonely
bound-state solution disappears asymptotically as one approaches the
conditions for the realization of spin and pseudospin symmetries.
\end{abstract}

\section{Introduction}

Over the years, relativistic potentials involving mixtures of vector and
scalar couplings has received attention in the literature. Heavy meson
spectra can be explained by solutions of the Dirac equation with a
convenient mixture of vector and scalar potentials (see, e.g., \cite{luc}).
The same can be said about the treatment of the nuclear phenomena describing
the influence of the nuclear medium on the nucleons \cite{ser}. Spin and
pseudospin symmetries are SU(2) symmetries of a Dirac equation with vector
and scalar potentials realized when the difference between the potentials,
or their sum, is a constant. The near realization of these symmetries may
explain degeneracies in some heavy meson spectra (spin symmetry) \cite{pag}-%
\cite{gin} or in single-particle energy levels in nuclei (pseudospin
symmetry) \cite{gin}-\cite{ps}, when these physical systems are described by
relativistic mean-field theories with scalar and vector potentials. When
these symmetries are realized, they decouple the upper and lower components
of the Dirac equation, so the energy spectrum does not depend on the
spinorial structure, being identical to the spectrum of a spinless particle
\cite{spin0}. Although the scalar potential finds many of their applications
in nuclear and particle physics, it could also simulate an effective mass in
solid state physics and so it could be useful for modelling transitions
between structures such a Josephson junctions \cite{bra}. In fact, there has
been a continuous interest for solving the Dirac equations in the
four-dimensional space-time as well as in lower dimensions for a variety of
potentials and couplings. A few recent works have been devoted to the
investigation of the solutions of the Dirac equation by assuming that the
vector potential has the same magnitude as the scalar potential \cite{gum}
whereas other works take a more general mixing \cite{asc1}. The
one-dimensional step potential is of certain interest to model the
transition between two structures. In solid state physics, for example, a
step-like potential which changes continuously over an interval whose
dimensions are of the order of the interatomic distances can be used to
model the average potential which holds the conduction electrons in metals.
Due to weak potentials, relativistic effects are considered \ to be small in
solid state physics, but the Dirac equation can give relativistic
corrections to the results obtained from the nonrelativistic equation. In
the presence of strong potentials, though, the Schr\"{o}dinger equation must
be replaced by their relativistic counterparts. The background of the kink
configuration of the $\phi ^{4}$ model \cite{raj} is of interest in quantum
field theory where topological classical backgrounds are responsible for
inducing a fractional fermion number on the vacuum. Models of these kinds,
known as kink models are obtained in quantum field theory as the continuum
limit of linear polymer models \cite{gol}. In a recent paper the complete
set of bound states of fermions in the presence of this sort of kink-like
smooth step potential has been addressed by considering a pseudoscalar
coupling in the Dirac equation \cite{asc}.

In the present work the scattering a fermion in the background of a sign
potential is considered with a general mixing of vector and scalar Lorentz
structures with the scalar coupling stronger than or equal to the vector
coupling. It is shown that a special unitary transformation preserving the
form of the current decouples the upper and lower components of the Dirac
spinor. Then the scattering problem is assessed under a Sturm-Liouville
perspective. The unique pole in the transmission amplitude is not related to
a proper bound state. Nevertheless, an isolated solution from the
Sturm-Liouville perspective is present. It is shown that, when the magnitude
of the scalar coupling exceeds the vector coupling, the fermion under a
strong potential can be trapped in a highly localized region without
manifestation of Klein's paradox. It is also shown that this curious lonely
bound-state solution disappears asymptotically as one approaches the
conditions for the realization of spin and pseudospin symmetries.

\section{Scalar and vector potentials in the Dirac equation}

The Dirac equation for a free fermion of rest mass $m$ reads%
\begin{equation}
\left( \gamma ^{\mu }p_{\mu }-Imc\right) \Psi =0  \label{d1}
\end{equation}%
where $p_{\mu }=i\hbar \partial _{\mu }$ is the momentum operator, $c$ is
the velocity of light, $I$ is the unit matrix and the square matrices $%
\gamma ^{\mu }$ satisfy the algebra $\{\gamma ^{\mu },\gamma ^{\nu
}\}=2Ig^{\mu \nu }$ in order to all the components of the spinor $\Psi $
satisfy the Klein-Gordon equation. In 1+1 dimensions $\Psi $ is a 2$\times $%
1 matrix and the metric tensor is $g^{\mu \nu }=$ diag$\left( 1,-1\right) $.
Eq. (\ref{d1}) can be written in the form
\begin{equation}
i\hbar \frac{\partial \Psi }{\partial t}=H_{0}\Psi  \label{b}
\end{equation}%
with the Hamiltonian given as $H_{0}=-i\gamma ^{5}cp_{1}+\gamma ^{0}mc^{2}$,
where $\gamma ^{5}=i\gamma ^{0}\gamma ^{1}$. It is worthwhile to note that
the matrices $\gamma ^{\mu }$ and $\gamma ^{5}$ are traceless, and the set $%
\{I$, $\gamma ^{0}$, $\gamma ^{1}$, $\gamma ^{5}\}$, where now $I$ is the 2$%
\times $2 unit matrix, is a complete set of linearly independent 2$\times $2
matrices. Requiring $\left( \gamma ^{\mu }\right) ^{\dag }=\gamma ^{0}\gamma
^{\mu }\gamma ^{0}$ and defining the adjoint spinor $\bar{\Psi}=\Psi
^{\dagger }\gamma ^{0}$, one finds the continuity equation $\partial _{\mu
}J^{\mu }=0$, where the conserved current is $J^{\mu }=c\bar{\Psi}\gamma
^{\mu }\Psi $. The positive-definite function $J^{0}/c=|\Psi |^{2}$, is
interpreted as a position probability density and its norm is a constant of
motion. This interpretation is completely satisfactory for single-particle
states \cite{tha}.

With the introduction of interactions, the Dirac equation can be written as%
\begin{equation}
\left( \gamma ^{\mu }p_{\mu }-Imc-V/c\right) \Psi =0  \label{d2}
\end{equation}%
and the current obeys the equation
\begin{equation}
\partial _{\mu }J^{\mu }=\frac{i}{\hbar }\bar{\Psi}\left( \gamma ^{0}V^{\dag
}\gamma ^{0}-V\right) \Psi  \label{j}
\end{equation}%
The current $J^{\mu }$ is still conserved provided $V^{\dag }=\gamma
^{0}V\gamma ^{0}$. The most general matrix potential preserving the current
conservation can be written in terms of well-defined Lorentz structures as
\begin{equation}
V=\gamma ^{\mu }A_{\mu }+IV_{s}+\gamma ^{5}V_{p}
\end{equation}%
We say that $A_{\mu }$, $V_{s}$ and $V_{p}$ are the vector, scalar and
pseudoscalar potentials, respectively, because the bilinear forms $\bar{\Psi}%
\gamma ^{\mu }\Psi $, $\bar{\Psi}I\Psi $ and $\bar{\Psi}\gamma ^{5}\Psi $
behave like vector, scalar and pseudoscalar quantities under a Lorentz
transformation, respectively. In this case the Hamiltonian can be written as%
\begin{equation}
H=-i\gamma ^{5}c\left( p_{1}+\frac{A_{1}}{c}\right) +IA_{0}+\gamma
^{0}\left( mc^{2}+V_{s}\right) +i\gamma ^{1}V_{p}
\end{equation}%
If the potentials are time independent one can write $\Psi \left( x,t\right)
=\psi \left( x\right) \exp \left( -iEt/\hbar \right) $ in such a way that
the time-independent Dirac equation becomes $H\psi =E\psi $. Meanwhile $%
J^{\mu }$ is time independent and $J^{1}$ is uniform. The space component of
the vector potential can be gauged away by defining a new spinor just
differing from the old by a phase factor so that we can consider $A_{1}=0$
without loss of generality.

Introducing the unitary operator

\begin{equation}
U(\theta )=\exp \left( -\frac{\theta }{2}\gamma ^{5}\right)  \label{2}
\end{equation}

\noindent where $\theta $ is a real quantity such that $0\leq \theta \leq
\pi $, one can write%
\begin{equation}
h\phi =E\phi  \label{h}
\end{equation}%
where $\phi =U\psi $, and $h=UHU^{-1}$ takes the form

\begin{equation}
h=-i\gamma ^{5}cp_{1}+IA_{0}+\gamma ^{0}\left[ \left( mc^{2}+V_{s}\right)
\cos \theta -V_{p}\sin \theta \right] +i\gamma ^{1}\left[ \left(
mc^{2}+V_{s}\right) \sin \theta +V_{p}\cos \theta \right]  \label{3}
\end{equation}%
It is instructive to note that the transformation preserves the form of the
current in such a way that $J^{\mu }=c\bar{\Phi}\gamma ^{\mu }\Phi $.

From now on, we do $V_{p}=0$ and use the representation $\gamma ^{0}=\sigma
_{3}$ and $\gamma ^{1}=i\sigma _{2}$ in such a way that $\gamma ^{5}=i\sigma
_{1}$. Here, $\sigma _{1}$, $\sigma _{2}$ and $\sigma _{3}$ stand for the
Pauli matrices.

The charge-conjugation operation is accomplished by the transformation $\psi
_{c}=\sigma _{1}\psi ^{\ast }$ followed by $A_{0}\rightarrow -A_{0}$, $%
V_{s}\rightarrow V_{s}$ and $E\rightarrow -E$. One sees that the charge
conjugation interchanges the roles of the upper and lower components of the
Dirac spinor. As a matter of fact, $A_{0}$ distinguishes fermions from
antifermions but $V_{s}$ does not, and so the spectrum is symmetrical about $%
E=0$ in the case of a pure scalar potential.

The two-component spinor $\psi $ can be written as $\psi ^{T}=\left( \psi
_{+}\,\psi _{-}\right) $, and if the potentials are small compared to $%
mc^{2} $ and $E\approx \pm mc^{2}$ one can write%
\begin{equation}
\psi _{\mp }\approx \pm \frac{p_{1}}{2mc}\,\psi _{\pm }  \label{nr1}
\end{equation}%
\begin{equation}
-\frac{\hbar ^{2}}{2m}\frac{d^{2}\psi _{\pm }}{dx^{2}}+\left( V_{s}\pm
A_{0}\right) \,\psi _{\pm }=\left( E\mp mc^{2}\right) \,\psi _{\pm }
\label{nr2}
\end{equation}%
Equation (\ref{nr1}) shows that $\psi _{\mp }$ is of order $v/c\ll 1$
relative to $\psi _{\pm }$, and equation (\ref{nr2}) shows that $\psi _{\pm
} $ obeys the Schr\"{o}dinger equation with the effective potential $%
V_{s}\pm A_{0}$ and energy equal to $E\mp mc^{2}$. This means that fermions
(antifermions), for weak potentials, are subject to the effective potential $%
V_{s}+A_{0}$ ($V_{s}-A_{0}$) with energy $E\approx \pm mc^{2}$. Therefore, a
mixed potential with $A_{0}=-V_{s}$ ($A_{0}=+V_{s}$) is associated with free
fermions (antifermions) in a nonrelativistic regime.

In terms of the upper and the lower components of the spinor $\phi $,
\noindent the Dirac equation decomposes into:
\begin{equation}
\hbar c\frac{d\phi _{\pm }}{dx}\pm \left( mc^{2}+V_{s}\right) \sin \theta
\,\phi _{\pm }=i\left[ E\pm \left( mc^{2}+V_{s}\right) \cos \theta -A_{0}%
\right] \phi _{\mp }  \label{eq1}
\end{equation}

\noindent Furthermore,%
\begin{equation}
\frac{J^{0}}{c}=|\phi _{+}|^{2}+|\phi _{-}|^{2},\quad \frac{J^{1}}{c}=2\text{%
Re}\left( \phi _{+}^{\ast }\phi _{-}\right)
\end{equation}

\noindent

\noindent Choosing
\begin{equation}
A_{0}=V_{s}\cos \theta  \label{5}
\end{equation}%
\noindent i.e., $|V_{s}|\geq $ $|A_{0}|$, one has
\begin{eqnarray}
\hbar c\frac{d\phi _{+}}{dx}+\left( mc^{2}+V_{s}\right) \sin \theta \,\phi
_{+} &=&i\left( E+mc^{2}\cos \theta \right) \phi _{-}  \label{6a} \\
&&  \notag \\
\hbar c\frac{d\phi _{-}}{dx}-\left( mc^{2}+V_{s}\right) \sin \theta \,\phi
_{-} &=&i\left[ E-\left( mc^{2}+2V_{s}\right) \cos \theta \right] \phi _{+}
\label{6b}
\end{eqnarray}

Note that $A_{0}$ changes its sign when the mixing angle $\theta $ goes from
$\pi /2-\varepsilon $ to $\pi /2+\varepsilon $. We now split two classes of
solutions depending on whether $E$ is different of or equal to $-mc^{2}\cos
\theta $:

\subsection{The class $E\neq -mc^{2}\cos \protect\theta $}

\noindent Using the expression for $\phi _{-}$ obtained from (\ref{6a}), viz.

\begin{equation}
\phi _{-}=\frac{-i}{E+mc^{2}\cos \theta }\left[ \hbar c\frac{d\phi _{+}}{dx}%
+\left( mc^{2}+V_{s}\right) \sin \theta \,\phi _{+}\right]  \label{7}
\end{equation}

\noindent one finds%
\begin{equation}
J^{1}=\frac{2\hbar c^{2}}{E+mc^{2}\cos \theta }\,\text{Im}\left( \phi
_{+}^{\ast }\frac{d\phi _{+}}{dx}\right)
\end{equation}%
Inserting (\ref{7}) in (\ref{6b}) one arrives at the following second-order
differential equation for $\phi _{+}$:
\begin{equation}
-\frac{\hbar ^{2}}{2}\frac{d^{2}\phi _{+}}{dx^{2}}+V_{\mathtt{eff}}\,\phi
_{+}=E_{\mathtt{eff}}\,\phi _{+}  \label{8}
\end{equation}%
where%
\begin{equation}
V_{\mathtt{eff}}=\frac{\sin ^{2}\theta }{2c^{2}}V_{s}^{2}+\frac{mc^{2}+E\cos
\theta }{c^{2}}V_{s}-\frac{\hbar \sin \theta }{2c}\frac{dV_{s}}{dx}
\label{v}
\end{equation}%
and%
\begin{equation}
E_{\mathtt{eff}}=\frac{E^{2}-m^{2}c^{4}}{2c^{2}}  \label{e}
\end{equation}

\noindent Therefore, the solution of the relativistic problem for this class
is mapped into a Sturm-Liouville problem for the upper component of the
Dirac spinor. In this way one can solve the Dirac problem for determining
the possible discrete or continuous eigenvalues of the system by recurring
to the solution of a Schr\"{o}dinger-like problem. For the case of a pure
scalar coupling ($E\neq 0$), it is also possible to write a second-order
differential equation for $\phi _{-}$ just differing from the equation for $%
\phi _{+}$ in the sign of the term involving $dV_{s}/dx$, namely,
\begin{equation}
-\frac{\hbar ^{2}}{2}\frac{d^{2}\phi _{\pm }}{dx^{2}}+\left( \frac{V_{s}^{2}%
}{2c^{2}}+mV_{s}\mp \frac{\hbar }{2c}\frac{dV_{s}}{dx}\right) \phi _{\pm
}=E_{\mathtt{eff}}\,\phi _{\pm }  \label{esc}
\end{equation}

\noindent This supersymmetric structure of the two-dimensional Dirac
equation with a pure scalar coupling has already been appreciated in the
literature \cite{coo}.

\subsection{The class $E=-mc^{2}\cos \protect\theta $}

Defining $v\left( x\right) =\int^{x}dy\,V_{s}\left( y\right) $, the
solutions for (\ref{6a})-(\ref{6b}) are%
\begin{eqnarray}
\phi _{+} &=&N_{+}  \notag \\
&&  \label{ii1} \\
\phi _{-} &=&N_{-}-2\frac{i}{\hbar c}N_{+}\left[ mc^{2}x+v\left( x\right) %
\right] \cos \theta  \notag
\end{eqnarray}%
for $\sin \theta =0$, and%
\begin{eqnarray}
\phi _{+} &=&N_{+}\exp \left\{ -\frac{\sin \theta }{\hbar c}\left[
mc^{2}x+v\left( x\right) \right] \right\}  \notag \\
&&  \label{ii2} \\
\phi _{-} &=&N_{-}\exp \left\{ +\frac{\sin \theta }{\hbar c}\left[
mc^{2}x+v\left( x\right) \right] \right\} +i\phi _{+}\cot \theta  \notag
\end{eqnarray}%
for $\sin \theta \neq 0$. \noindent $N_{+}$ and $N_{-}$ are normalization
constants. It is instructive to note that there is no solution for
scattering states. Both set of solutions present a space component for the
current equal to $J^{1}=2c\text{Re}\left( N_{+}^{\ast }N_{-}\right) $ and a
bound-state solution demands $N_{+}=0$ or $N_{-}=0$, because $\phi _{+}$ and
$\phi _{-}$ are square-integrable functions vanishing as $|x|\rightarrow
\infty $. There is no bound-state solution for $\sin \theta =0$, and for $%
\sin \theta \neq 0$ the existence of a bound state solution is possible only
if $v(x)$ has a distinctive leading asymptotic behaviour \cite{hot}. Note
also that%
\begin{equation}
\phi _{\pm }=N_{\pm }\exp \left\{ \mp \frac{1}{\hbar c}\left[
mc^{2}x+v\left( x\right) \right] \right\}  \label{sc1}
\end{equation}%
in the case of a pure scalar coupling ($E=0$), so that either $\phi _{+}=0$
or $\phi _{-}=0$.\noindent

\section{The sign potential}

Consider now the potential%
\begin{equation}
V_{s}=v_{0}\,\text{sgn}\left( x\right)
\end{equation}%
where sgn$\left( x\right) =x/|x|$ ($x\neq 0$) is the sign function, so that $%
v\left( x\right) =v_{0}|x|$. Our problem is to solve the set of equations (%
\ref{6a})-(\ref{6b}) for $\phi $ and to determine the allowed energies.

\subsection{The case $E\neq -mc^{2}\cos \protect\theta $}

For our model, recalling (\ref{7}) and (\ref{v}), one finds%
\begin{equation}
\phi _{-}=\frac{-i}{E+mc^{2}\cos \theta }\left\{ \hbar c\frac{d\phi _{+}}{dx}%
+\left[ mc^{2}+v_{0}\,\text{sgn}\left( x\right) \right] \sin \theta \,\phi
_{+}\right\}
\end{equation}%
\begin{equation}
V_{\mathtt{eff}}=\frac{v_{0}^{2}\sin ^{2}\theta }{2c^{2}}+\frac{mc^{2}+E\cos
\theta }{c^{2}}\,v_{0}\,\text{sgn}\left( x\right) -\frac{\hbar v_{0}\sin
\theta }{c}\,\delta \left( x\right)  \label{pot}
\end{equation}%
For $\sin \theta =0$, the \textquotedblleft effective
potential\textquotedblright\ is an ascendant (a descendant) step if $%
(mc^{2}+E\cos \theta )v_{0}>0$ ($(mc^{2}+E\cos \theta )v_{0}<0$). For $\sin
\theta \neq 0$, the \textquotedblleft effective potential\textquotedblright\
includes an attractive (a repulsive) delta function at the origin if $%
v_{0}>0 $ ($v_{0}<0$). Therefore, we expect scattering states in all the
circumstances. Nevertheless, bound-state solutions should not be expected
for $\sin \theta \neq 0$ and $v_{0}<0$.

We demand that $\phi _{+}$ be continuous at $x=0$, that is%
\begin{equation}
\left. \phi _{+}\right\vert _{x=0_{+}}-\left. \phi _{+}\right\vert
_{x=0_{-}}=0  \label{cont}
\end{equation}%
Otherwise, the differential equation for $\phi _{+}$ would contain the
derivative of a $\delta $-function. Effects on $d\phi _{+}/dx$ in the
neighbourhood of $\ x=0$ can be evaluated by integrating the differential
equation for $\phi _{+}$ from $-\varepsilon $ to $+\varepsilon $ and taking
the limit $\varepsilon \rightarrow 0$. The connection formula between $d\phi
_{+}/dx$ at the right and $d\phi _{+}/dx$ at the left can be summarized as%
\begin{equation}
\left. \frac{d\phi _{+}}{dx}\right\vert _{x=0_{+}}-\left. \frac{d\phi _{+}}{%
dx}\right\vert _{x=0_{-}}=-\frac{2v_{0}\sin \theta }{\hbar c}\,\phi
_{+}\left( 0\right)   \label{discont}
\end{equation}%
One can also see that, despite the discontinuity of $d\phi _{+}/dx$ for $%
\sin \theta \neq 0$, $J^{\mu }$ is a continuous function.

\subsubsection{Scattering states}

We turn our attention to scattering states for fermions coming from the
left. Then, $\phi $ for $x<0$ describes an incident wave moving to the right
and a reflected wave moving to the left, and $\phi $ for $x>0$ describes a
transmitted wave moving to the right or an evanescent wave. The upper
components for scattering states are written as
\begin{equation}
\phi _{+}=\left\{
\begin{array}{cc}
Ae^{+ik_{-}x}+Be^{-ik_{-}x} & \text{for\quad }x<0 \\
&  \\
C_{\pm }e^{\pm ik_{+}x} & \text{for\quad }x>0%
\end{array}%
\right.  \label{phi}
\end{equation}%
where

\begin{equation}
\hbar ck_{\pm }=\sqrt{\left( E\mp v_{0}\cos \theta \right) ^{2}-\left(
mc^{2}\pm v_{0}\right) ^{2}}
\end{equation}%
Note that $k_{+}$ is a real number for a progressive wave and an imaginary
number an evanescent wave ($k_{-}$ is a real number). Therefore,%
\begin{equation}
J^{1}\left( x<0\right) =\frac{2\hbar c^{2}k_{-}}{E+mc^{2}\cos \theta }\left(
|A|^{2}-|B|^{2}\right)
\end{equation}%
and
\begin{equation}
J^{1}\left( x>0\right) =\pm \,\frac{2\hbar c^{2}\text{Re}\,k_{+}}{%
E+mc^{2}\cos \theta }\,|C_{\pm }|^{2}
\end{equation}%
In fact, if $E>-mc^{2}\cos \theta $, then $Ae^{+ik_{-}x}$ ($Be^{-ik_{-}x}$)
will describe the incident (reflected) wave, and $C_{-}=0$. On the other
hand, if $E<-mc^{2}\cos \theta $, then $Be^{-ik_{-}x}$ ($Ae^{+ik_{-}x}$)
will describe the incident (reflected) wave, and $C_{+}=0$.

With the connection formulas (\ref{cont}) and (\ref{discont}), and omitting
the algebraic details, we state the solution for the transmission amplitudes%
\begin{eqnarray}
\frac{C_{+}}{A} &=&\frac{2k_{-}}{k_{-}+k_{+}-i2v_{0}\sin \theta /\left(
\hbar c\right) }  \notag \\
&& \\
\frac{C_{-}}{B} &=&\frac{2k_{-}}{k_{-}+k_{+}+i2v_{0}\sin \theta /\left(
\hbar c\right) }  \notag
\end{eqnarray}%
To determine the transmission coefficient we use the current densities $%
J^{1}\left( x<0\right) $ and $J^{1}\left( x>0\right) $. The $x$-independent
space component of the current allows us to define the transmission
coefficient as%
\begin{equation}
T=\frac{\text{Re}\,k_{+}}{k_{-}}\frac{|C_{+}|^{2}}{|A|^{2}}=\frac{\text{Re}%
\,k_{+}}{k_{-}}\frac{|C_{-}|^{2}}{|B|^{2}}
\end{equation}%
so that%
\begin{equation}
T=\frac{4k_{-}\text{Re}\,k_{+}}{\left( k_{-}+k_{+}\right) ^{2}+\left( \frac{%
2v_{0}\sin \theta }{\hbar c}\right) ^{2}}
\end{equation}%
taking no regard if $E>-mc^{2}\cos \theta $ or $E<-mc^{2}\cos \theta $.
Nevertheless, scattering states are possible only if $|E+v_{0}\cos \theta
|>|mc^{2}-v_{0}|$ because $k_{-}$ is a real number, and there is a
transmitted wave only if $|E-v_{0}\cos \theta |>|mc^{2}+v_{0}|$. For $\theta
=0$ and $v_{0}=mc^{2}$, the transmission coefficient as a function of $E$ is
illustrated in Figure 1. As $|E|\rightarrow \infty $, $T\rightarrow 1$ as it
should. For $E>-mc^{2}\cos \theta $, this is a profile typical for the
nonrelativistic scattering in a step potential or in a delta potential. The
same profile is observed for other values of $\theta $ and $v_{0}$. Figure 2
shows the transmission coefficient as a function of the mixing angle. Those
intriguing results are readily explained by observing that the effective
potential presents an ascendant (descendant) step for small (large) values
of $\theta $. The transmission coefficient vanishes for enough small mixing
angles and energies because the effective energy is smaller than the height
of the effective step potential. For $|v_{0}|>mc^{2}$, the absence of
scattering for enough large mixing angles and enough small energies occurs
because the effective energy is smaller than the effective step potential in
the region of incidence ($x<0$).

\subsubsection{Bound states}

The possibility of bound states requires a solution given by (\ref{phi})
with $k_{\pm }=i|k_{\pm }|$ and $A=0$, or $k_{\pm }=-i|k_{\pm }|$ and $B=0$,
to obtain a square-integrable $\phi _{+}$. This means that $|E\pm v_{0}\cos
\theta |<|mc^{2}\mp v_{0}|$. On the other hand, if one considers the
transmission amplitude as a function of the complex variables $k_{\pm }$ one
sees that for $k_{\pm }>0$ one obtains the scattering states whereas the
bound states would be obtained by the poles lying along the imaginary axis
of the complex $k$-plane. These poles are given by the equation
\begin{equation}
|k_{-}|+|k_{+}|=\frac{2v_{0}\sin \theta }{\hbar c}  \label{cq}
\end{equation}%
Equation (\ref{cq}) is the quantization condition. Because $|k_{-}|+|k_{+}|$
is a positive quantity, one concludes the bound states are impossible for $%
v_{0}<0$ or $\sin \theta =0$, as has been anticipated from the qualitative
discussions. In fact, squaring (\ref{cq}) results in the form of a
second-degree algebraic equation%
\begin{equation}
E^{2}+2mc^{2}\cos \theta \,E+m^{2}c^{4}\cos ^{2}\theta =0  \label{cq1}
\end{equation}%
which presents just one solution: $E=-mc^{2}\cos \theta $. Evidently, it is
not a proper solution of the problem.

\subsection{The case $E=-mc^{2}\cos \protect\theta $}

As commented before, there is no solution for $\sin \theta =0$, and the
normalizable solution for $\sin \theta \neq 0$ requires $|v_{0}|>mc^{2}$:%
\begin{equation}
\phi =\left(
\begin{array}{c}
1 \\
i\cot \theta
\end{array}%
\right) N_{>}\exp \left\{ -\frac{\sin \theta }{\hbar c}\left[ |v_{0}|+mc^{2}%
\text{sgn}\left( x\right) \right] |x|\right\}   \label{s1}
\end{equation}%
for $v_{0}>mc^{2}$, and%
\begin{equation}
\phi =\left(
\begin{array}{c}
0 \\
1%
\end{array}%
\right) N_{<}\exp \left\{ -\frac{\sin \theta }{\hbar c}\left[ |v_{0}|-mc^{2}%
\text{sgn}\left( x\right) \right] |x|\right\}   \label{s2}
\end{equation}%
for $v_{0}<mc^{2}$. Here,
\begin{equation}
N_{>}=N_{<}\sin \theta =\sin \theta \sqrt{\frac{\sin \theta }{\hbar c}\frac{%
v_{0}^{2}-m^{2}c^{4}}{|v_{0}|}}  \label{ene}
\end{equation}%
From (\ref{s1}) and (\ref{s2}), one readily finds the position probability
density to be%
\begin{equation}
|\phi |^{2}=\frac{\sin \theta }{\hbar c}\frac{v_{0}^{2}-m^{2}c^{4}}{|v_{0}|}%
\exp \left\{ -\frac{2\sin \theta }{\hbar c}\left[ |v_{0}|+mc^{2}\text{sgn}%
\left( v_{0}\right) \text{sgn}\left( x\right) \right] |x|\right\}
\label{den}
\end{equation}%
The position probability density is graphed in Fig. 3 for two different signs of $v_{0}$. The expectation value of $x$ given by%
\begin{equation}
<x>=-\text{sgn}\left( v_{0}\right) \frac{\hbar c}{\sin \theta }\frac{mc^{2}}{%
v_{0}^{2}-m^{2}c^{4}}  \label{espec}
\end{equation}%
Therefore, the fermion tends to concentrate at the left (right) region when $%
v_{0}>0$ ($v_{0}<0$), and tends to avoid the origin more and more as $\sin
\theta $ decreases. The fermion is confined within an interval $\Delta x=%
\sqrt{<x^{2}>-<x>^{2}}$ given by%
\begin{equation}
\Delta x=\frac{\hbar c}{\sqrt{2}\sin \theta }\frac{\sqrt{v_{0}^{2}+m^{2}c^{4}%
}}{v_{0}^{2}-m^{2}c^{4}}  \label{dx}
\end{equation}%
One can see that the best localization occurs for a pure scalar coupling. In
fact, the fermion becomes delocalized as $\sin \theta $ decreases. If $%
\Delta x$ shrinks with rising $|v_{0}|$ or $\sin \theta $ then $\Delta p$
(uncertainty in the momentum) will swell, in consonance with the Heisenberg
uncertainty relation. Nevertheless, the maximum uncertainty in the momentum
is given by $mc$ requiring that is impossible to localize a fermion in a
region of space less than half of its Compton wavelength (see, for example,
\cite{gre}). Nevertheless, if one defines an effective mass as $m_{\mathtt{%
eff}}=m\sqrt{1+\left( v_{0}/mc^{2}\right) ^{2}}$ and an effective Compton
wavelength $\lambda _{\mathtt{eff}}=\hbar /\left( m_{\mathtt{eff}}c\right) $%
, one will find%
\begin{equation}
\Delta x=\frac{\lambda _{\mathtt{eff}}}{\sqrt{2}\sin \theta }\frac{%
v_{0}^{2}+m^{2}c^{4}}{v_{0}^{2}-m^{2}c^{4}}  \label{dx1}
\end{equation}%
It follows that the high localization of fermions, related to high values of
$v_{0}$, never menaces the single-particle interpretation of the Dirac
theory even if the fermion is massless. This fact is convincing because the
scalar coupling exceeds the vector coupling, and so the conditions for
Klein's paradox are never reached.

\section{Final remarks}

We have assessed the stationary states of a fermion under the influence of a
sign potential for a special mixing of scalar and vector couplings. A
special unitary transformation allowed to decouple the upper and lower
components of the Dirac spinor and to assess the scattering problem under a
Sturm-Liouville perspective. An isolated solution appears when the vector
coupling and the scalar coupling have different magnitudes showing that the
fermion under a strong potential can be trapped in a highly localized region
without the possibility of pair production. It was also shown that this
curious lonely bound-state solution disappears asymptotically as one
approaches the conditions for the realization of spin and pseudospin
symmetries.

\bigskip

\bigskip

\bigskip

\noindent \textbf{Acknowledgments}

This work was supported in part by means of funds provided by CNPq.

\newpage

\begin{figure}[th]
\begin{center}
\includegraphics[width=10cm]{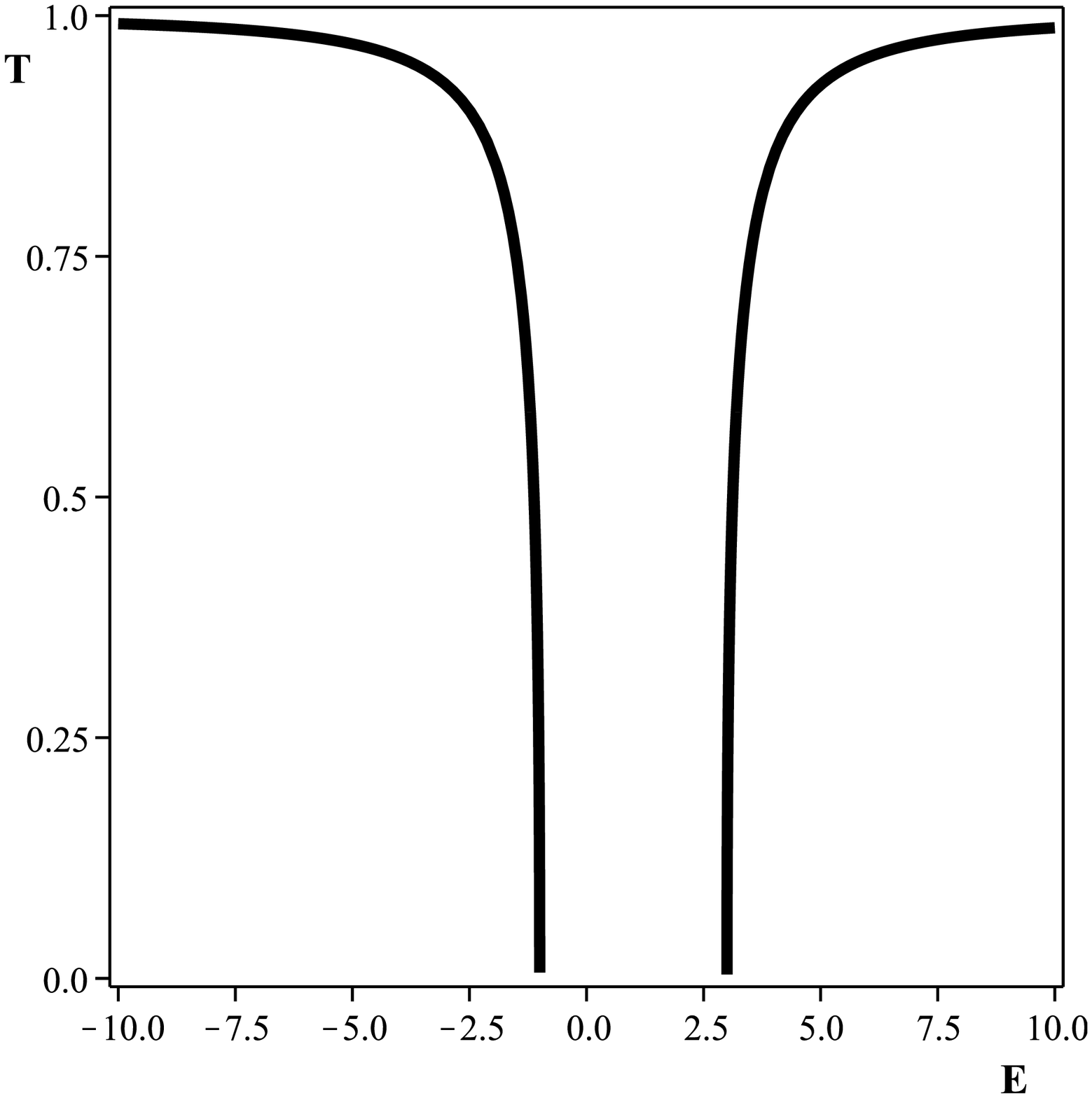} \label{fig:Fig1}
\end{center}
\par
\vspace*{-0.1cm}
\caption{ Transmission coefficient as a function of energy for $\protect%
\theta =0$ and $v_{0}=1$ ($\hbar =m=c=1$).}
\end{figure}

\begin{figure}[th]
\begin{center}
\subfigure[$v_{0}=0.9$]{
\includegraphics[width=10cm]{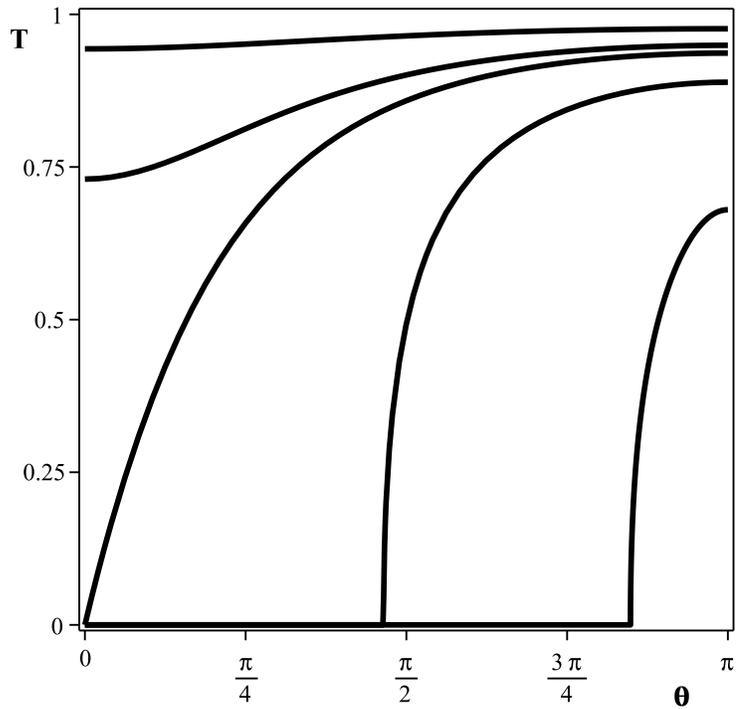}
\label{fig:Fig2a}
}
\par
\subfigure[$v_{0}=1.1$]{
\includegraphics[width=10cm]{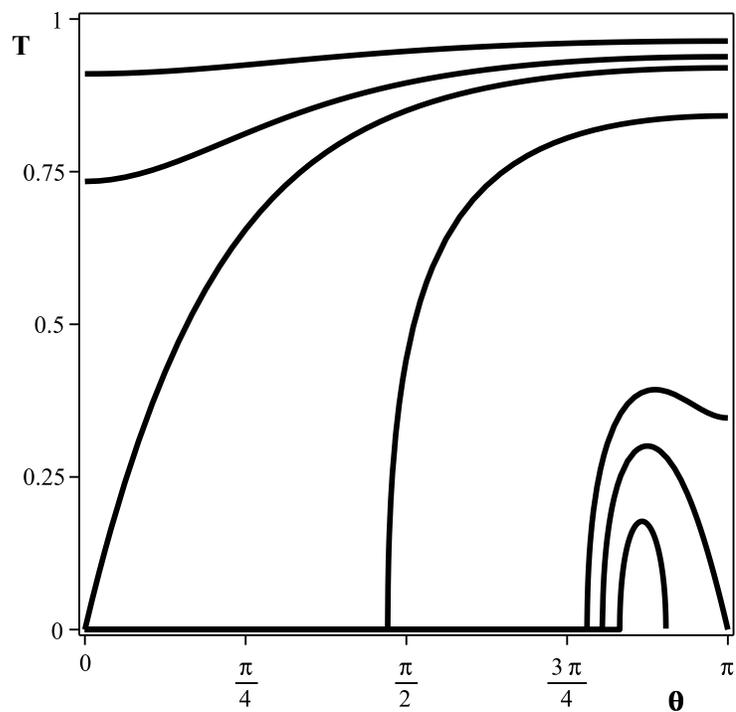}
\label{fig:Fig2b}
}
\end{center}
\caption{Transmission coefficient as a function of $\protect\theta $ .
Higher coefficients correspond to higher energies. The most inferior curve
corresponds to $E$=1.1 in (a) and to 1.15 in (b) ($\hbar =m=c=1$). A more
detailed description is given in the text.}
\end{figure}

\begin{figure}[th]
\begin{center}
\includegraphics[width=10cm]{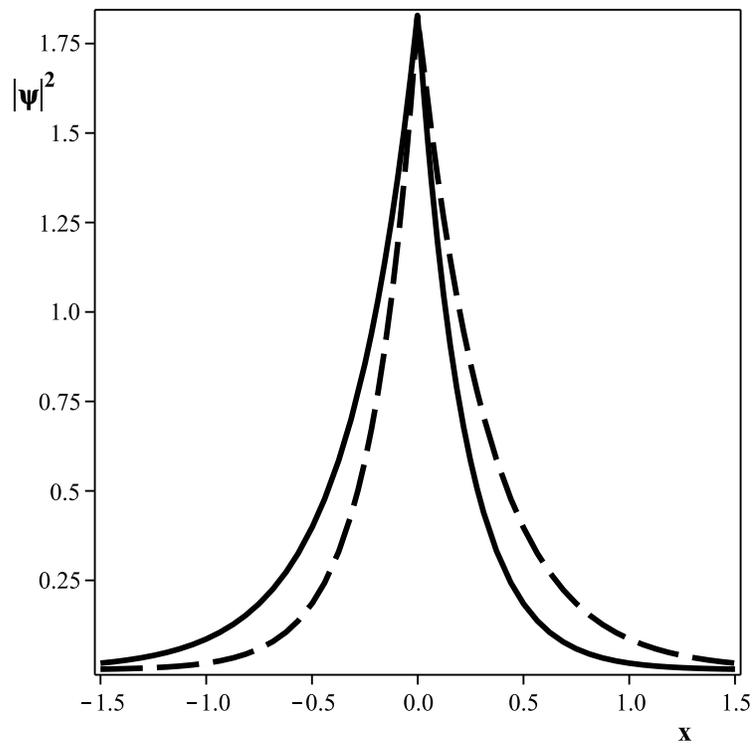} \label{fig:Fig3}
\end{center}
\par
\vspace*{-0.1cm}
\caption{ Probability density for $\protect\theta = \protect\pi /8$. The
continuous (dashed) line for $v_{0}=+5 \; (-5)$ ($\hbar =m=c=1$).}
\end{figure}

\end{document}